# Emotionally Vulnerable Subtype of Internet Gaming Disorder: Measuring and Exploring the Pathology of Problematic Generative AI Use

Haocan Sun [a,b,†], Di Wu [a†], Weizi Liu [c], Guoming Yu [a*], Mike Yao [b*]

[a] *School of Journalism and Communication, Beijing Normal University, Beijing, China, 100875*

[b] *Institute of Communications Research, College of Media, University of Illinois Urbana-Champaign, Urbana, IL, USA, 61801*

[c] *Bob Schieffer College of Communication, Texas Christian University, Fort Worth, TX, USA, 76129*

[†] Equal contributor

*Corresponding Author:

## Abstract

Concerns over the potential over-pathologization of generative AI (GenAI) use and the lack of conceptual clarity surrounding GenAI addiction call for empirical tools and theoretical refinement. This study developed and validated the PUGenAIS-9 (Problematic Use of Generative Artificial Intelligence Scale-9 items) and examined whether PUGenAIS reflects addiction-like patterns under the Internet Gaming Disorder (IGD) framework. Using samples from China and the United States (N = 1,508), we conducted confirmatory factor analysis and identified a robust 31-item structure across nine IGD-based dimensions. We then derived the PUGenAIS-9 by selecting the highest-loading items from each dimension and validated its structure in an independent sample (N = 1,426). Measurement invariance tests confirmed its stability across nationality and gender. Person-centered (latent profile analysis) and variable-centered (network analysis) approaches revealed a 5–10% prevalence rate, a symptom network structure similar to IGD, and predictive factors related to psychological distress and functional impairment. These findings indicate that PUGenAI shares features of the emotionally vulnerable subtype of IGD rather than the competence-based type. These results support using PUGenAIS-9 to identify problematic GenAI use and show the need to rethink digital addiction with an ICD (infrastructures, content, and device) model. This keeps addiction research responsive to new media while avoiding over-pathologizing.

## Introduction

Large language model-based Generative AI (GenAI) encourages people to proactively and consciously use interpersonal communication scripts in Human–AI interaction (Kühne & Peter, 2023). Through dynamic engagement with the environment, GenAI fosters synthetic relationships and generates human-level outputs (Starke et al., 2024). However, problematic use of GenAI (PUGenAI) has begun to manifest in increasingly tangible and alarming ways. Recent reports have documented severe psychosocial harms arising from emotional interactions with GenAI, including cases of self-harm and mortality among vulnerable users (Duffy, 2024; Horwitz, 2025). Beyond these incidents, broader patterns indicate a growing dependence on GenAI. People increasingly treat GenAI as pseudo-colleagues, relying on it not only for task execution but also for emotional support. Studies reveal that 72 % adolescents treat chatbots as friends and advisers to accompany them (Robb & Mann, 2025).

However, the use of GenAI companions brings substantial risks. ChatGPT, which is accessible to children without age restrictions or parental controls, can produce harmful content within minutes of account registration (Center for Countering Digital Hate, 2025). 34% of AI companion users report feeling uncomfortable with something the companion has said or done (Robb & Mann, 2025). Latest analyses of millions of ChatGPT interactions combined with a four-week trial of nearly 1,000 users found that heavier chatbot usage correlates with elevated loneliness and emotional dependence and reduced offline social engagement (Phang et al., 2025). This trend raises concern that PUGenAI may already be emerging as a meaningful phenomenon, warranting theoretical and measurement attention.

So far, the term *AI addiction* has appeared in recent discourse. While these studies offer insights into its definition and encourage exploration of the underlying mechanisms, they often suffer from conceptual insularity (Ciudad-Fernández et al., 2025). The primary challenge is the establishment of consistent terminology and valid measurement instruments.

Emerging evidence suggests that frequent use of GenAI can lead to excessive reliance, loss of control, and affect regulation through the system (Yu et al., 2024). These behaviors resemble core symptoms of behavioral addiction, including impaired control, tolerance, withdrawal, and functional impairment (Kooli et al., 2025). While some scholars argue for the legitimacy of conceptualizing GenAI addiction as a new form of technology-related disorder, others caution against over-pathologization, warning that it may trivialize the clinical severity of addiction (Ciudad-Fernández et al., 2025). Therefore, the study firstly aims to explore whether PUGenAI shows a tendency towards behavioral addiction.

Exploring the pathological nature of PUGenAI needs to rely on precise measurements. Therefore, before unfolding the internal structure of PUGenAI, it is essential to first develop and validate an instrument of PUGenAI. Yet at least four scales have claimed to measure GenAI addiction (Ciudad-Fernández et al., 2025). However, these instruments have several flaws. First, most adopt a define-and-create approach, assuming that GenAI use necessarily leads to addiction, which risks over-pathologization. Second, none of them has undergone rigorous development procedures. They often rely on inadequate sample sizes, lack items derived from systematic qualitative inquiry, fail to separate development and validation datasets, and do not test measurement invariance. Moreover, they ultimately remain at the level of definition without further exploring the traits of addiction. To date, we still lack a clear understanding of the distinctive features of PUGenAI and a reasonable basis for inferring whether it leads to addiction (Ciudad-Fernández et al., 2025).

In response to recent calls for diagnostic criteria of GenAI addiction (Kooli et al., 2025) and cautions against hastily labeling GenAI use as addiction (Ciudad-Fernández et al., 2025), the present study adopts a cautious and data-driven approach (see Figure 1) through a multi-phase research program. Phase 1 included three sequential studies that together comprise the scale-development process. Study 1 focused on item generation and content validation: we built an initial item pool based on DSM-5 IGD, existed GenAI problematic use (addiction) measures, and insights from qualitative interviews to ensure coverage of GenAI's unique relational and competence-based dependence mechanism. Study 2 fielded this item pool in a large-scale cross-national survey in China and the U.S. (Wave 1) and conducted a confirmatory factor analysis (CFA) to replicate the latent structure of IGD. In study 3, we conducted a second large-scale survey with an independent validation sample (Wave 2) and validated the PUGenAI short form's psychometric properties, including: (a) factor structure and internal consistency, (b) convergent validity, (c) criterion-related validity, and (d) measurement invariance across nationalities and gender.

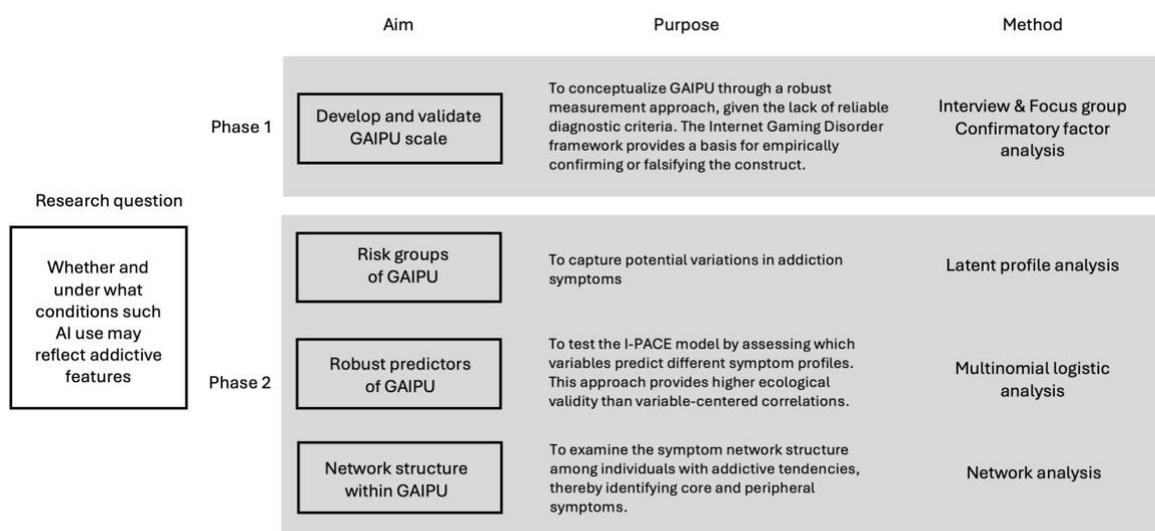

Figure 1. Framework of the Study

Phase 2 applied a complementary, variable- and person-centered lens to explore the internal structure of PUGenAI in wave 2 data. Specifically, to examine whether PUGenAI manifests features consistent with behavioral addiction, we conducted latent profile analysis and network analysis to explore (a) prevalence patterns, (b) symptom network structure, and (c) predictive factors.

Before exploring whether PUGenAI reflects a trait of addiction, we return to three theoretical foundational questions. First, what is the origin and the meaning of the word *addiction*? Second, why does PUGenAI have the potential to foster dependency? Third, how should PUGenAI be positioned within the landscape of digital addiction?

## literature review

### *Addiction, Problematic Use, Disorder, or Dependence*

The conceptual boundaries among addiction, problematic use, disorder, and dependence remain contested across psychiatry, clinical psychology, and media studies (Figure 2). *Addiction* is defined as a chronic and treatable medical condition. It arises from complex interactions among neural circuits, genetic predispositions, environmental factors, and individual life experiences (American Society of Addiction Medicine, 2019). However, despite its conceptual breadth, the term *addiction* is rarely used in the formal DSM-5. Scholars generally classify addiction into two categories: substance addiction (e.g., drug use) and behavioral addiction (e.g., internet gaming or gambling, see DSM-5). Kuss & Griffiths (2017) argued that behavioral and substance addictions share neurocognitive mechanisms. However, others have maintained that behavioral addictions are not directly comparable to substance-based disorders (Kardefelt-Winther et al., 2017; Panova & Carbonell, 2018). Therefore, critics warned that the inclusion of IGD in the DSM-5 risks over-pathologizing (Billieux et al., 2015; van Rooij et al., 2018).

| Term | Definition | Unresolved Issues | Using Context | Compare With Addiction |
|---|---|---|---|---|
| Addiction | Addiction is rarely used in substance use, but behavioral addiction is defined as a repeated behavior that causes significant, functionally impairing harm or distress, persists over time, and is not voluntarily reduced. | The conceptual boundaries of addiction remain ambiguous and often stigmatized. Lacking useful and valid symptoms in behavioral addiction research. | Behavior and substance addiction. behavioral addiction requires two components: (a) significant functional impairment or distress directly caused by the behavior, and (b) persistence over time. | |
| Problematic Use | Excessive but non-disordered behavior | Conceptually ambiguous but frequently used due to its non-diagnostic nature. | Preferred term in media and psychological research due to lower pathologization risk | Conceptually broader and more inclusive than addiction |
| Dependence | Highlights physiological adaptation such as tolerance and withdrawal to a substance. | Often conflated with behavioral disorders, but originally referred to pharmacological adaptation, including tolerance and withdrawal. | Applies only to substances; DSM-5 integrates *abuse* and *dependence* into a single continuum—substance use disorder | Adopted in DSM-III/DSM-III-R instead of "addiction" due to stigma concerns overlaps with Addiction in older models |
| Disorder | Requires clinically significant impairment, distress, or physiological symptoms based on diagnostic criteria. | Requires clinically significant impairment, distress, or physiological symptoms based on diagnostic criteria, yet the advancement of behavioral disorder diagnostics has lagged behind technological developments. | Requires meeting established clinical diagnostic criteria. | Clinical threshold; includes some Addictions, excludes many Problematic Uses |

*Note.* Adapted from prior frameworks (Alavi et al., 2012; Andreassen, 2015; Billieux et al., 2015, 2017; Nathan, 1991; O'Brien, 2011).

Figure 2. Conceptual Terms Commonly Used in Digital Addiction Research

Moreover, the term *addiction* has been deliberately omitted from the official nomenclature of DSM-5's substance use disorders due to definitional ambiguity and its stigmatizing connotation (American Psychiatric Association, 2013). In recent decades, the term has proliferated in public discourse and academic literature, often serving as a metaphor for overuse. In response to these concerns, researchers prefer the term *problematic use*, which captures patterns of maladaptive engagement without presuming clinical pathology (Billieux et al., 2017; Billieux et al., 2015), allowing for a more nuanced examination of behaviors that are excessive or harmful but may not meet the diagnostic criteria for a mental disorder.

The terms *abuse* and *dependence* appeared early in the clinical vocabulary of substance use. DSM-3 (Third Edition) distinguished abuse from dependence based on physiological criteria such as tolerance and withdrawal, leaving *abuse* as a residual category with limited diagnostic clarity (Nathan, 1991). *Dependence* was initially favored as a pharmacological term, describing physiological adaptation, including the development of tolerance and the experience of withdrawal symptoms (Widiger & Smith, 1994). However, its use as a substitute for *addiction* sparked controversy. In DSM-3-R (third edition, revised version), *dependence* was chosen over *addiction* by a narrow vote, due to concerns over stigma (O'Brien, 2011). However, clinicians strongly objected to this choice, noting that this conflation risked misrepresenting compulsive, pathological drug-seeking behavior (Horowitz & Taylor, 2023) To resolve this, DSM-V eliminated the abuse-dependence split by construct a continuum and adopted *Substance Use Disorder* as a diagnosis with graded severity, citing insufficient empirical evidence supporting an intermediate stage between casual use and full-blown addiction (American Psychiatric Association et al., 2013). While DSM-V includes disorders that reflect addiction, not all behavioral addictive qualify as disorders (O'Brien, 2011). Addiction is encompassed within the broader category of disorders, but is not synonymous with them. Therefore, since the appropriate clinical label for the identified latent construct remains uncertain, we do not assume a priori that PUGenAI qualifies as behavioral addiction, and provisionally call it problematic use.

### Why GenAI Use Might Be Dependence: Tool vs. Social Actor

How should PUGenAI be categorized? To label a behavior as addictive, there must be convincing evidence of negative consequences, impaired control, psychological distress, and functional impairment (Billieux et al., 2017; Kardefelt-Winther et al., 2017). Although studies have suggested that GenAI may lead to addictive behaviors (Kooli et al., 2025) and compulsive ChatGPT use has been linked to anxiety, burnout, and sleep disturbances (Duong, 2024), there is limited empirical evidence. Unlike gambling, or game disorder characterized by pleasure-seeking and reward-driven behaviors, GenAI's core utility lies in different ways. PUGenAI often involves a more immersive, agentic, and psychologically engaging process than previous forms of media interaction (Abbas et al., 2024). To understand the mechanisms underlying potential addiction to GenAI, scholars may adopt two perspectives: GenAI as a functional tool and a social actor (Choung et al., 2023).

From the instrumental perspective, GenAI operates as an intelligent assistant that enhances efficiency and supports cognitive outsourcing. By compensating for human limitations, it can improve decision accuracy and task performance (Klingbeil et al., 2024). Empirical studies have shown that using GenAI for writing tasks can increase productivity by 40% and improve output quality by 18% (Noy & Zhang, 2023). GenAI can also assist with everyday tasks such as planning and information retrieval (Ye et al., 2025). However, it also fosters excessive reliance, reduces critical thinking, and leads to habitual overuse (Yankouskaya et al., 2024). However, from a psychiatric standpoint, instrumental overuse alone does not qualify as addiction. For example, excessive reliance on smartphones has not been classified as a mental disorder in DSM-5. No form of clinically recognized addiction has been attributed to mere functional dependence. Therefore, a functional-dependence model, in isolation, cannot sufficiently justify the pathologization of GenAI use.

In contrast to functional dependence, emotional reliance on GenAI may carry greater addictive potential. From the perspective of Computers as Social Actors (CASA, Reeves & Nass, 1996), using GenAI forms synthetic relationships, fostering technologically mediated emotionally companionship (Starke et al., 2024). The increasing quality and frequency of social cues generated by AI amplify this tendency (Xu et al., 2022). Research shows that loneliness, trust, and the perception of human-likeness significantly predict both frequent use and psychological dependency on conversational agents (Xie et al., 2023). Yankouskaya et al. (2024) highlight how real-time feedback, personalized responses, and simulated social presence may jointly promote compulsive usage patterns.

*Positioning PUGenAI Among Digital Problematic Use*

PUGenAI shares conceptual ground with several established forms of digital problematic use, yet its underlying mechanisms reveal important distinctions. Existing forms of digital addiction can be broadly classified into three levels: device, infrastructure, and content. At the device or technological infrastructure level, smartphone problematic use, computer problematic use, and broader internet addiction emphasize the enabling medium rather than specifying the content or type of interaction (Charlton, 2002; Sun & Tang, 2025). In these cases, social function often plays a more central role in dependence than informational, leisure, and entertainment functions (Li & Chung, 2006). At the content level, IGD and social media disorder (SMD, or called social network site addiction) involve engagement with an activity such as gameplay or social networking (Andreassen, 2015; Gomez et al., 2022), whereas PUGenAI encompasses diverse functions. GenAI not only enables text and image generation (Anantrasirichai & Bull, 2022) and decision-making support (Lahat et al., 2024) but also increasingly exhibits anthropomorphic qualities, contextual awareness, and personalization. These features provide users with immediate feedback as well as cognitive and emotional support during interaction (Kocoń et al., 2023; Y. Li et al., 2024), including forms of emotional companionship (Pani et al., 2024; Skjuve et al., 2021).

In our view, GenAI systems can be conceptualized as a specific form of content, comparable to SMD and IGD, as they afford users a high degree of freedom to interact within an application, whether with AI agents, software programs, or other humans. The underlying technology, large language models, resembles the internet in that it constitutes a form of technological infrastructure accessible across different devices. By integrating multiple dependency pathways, including functional reliance, emotional attachment, and social surrogacy (Starke et al., 2024), PUGenAI not only reproduces the addictive dynamics observed in IGD and SMD (Van Den Eijnden et al., 2016) but also generates patterns of sustained engagement that existing frameworks may not fully capture.

Recent studies have offered some cues for positioning PUGenAI. Among various motivations for AI use, escape and social motivations mediate the relationship between mental health challenges and AI dependence, whereas entertainment and instrumental motivations do not (Huang et al., 2024). This pattern echoes the *emotionally vulnerable* subtype of IGD, characterized by high levels of internalizing symptoms such as anxiety and depression, often rooted in adverse early-life experiences (Lee et al., 2017). For individuals in this profile, internet gaming functions less as entertainment and more as an emotional escape or mood regulation mechanism. Empirical studies have reported that nearly one-third of role-playing gamers use gaming to avoid negative affect or psychological distress (Hussain & Griffiths, 2009). Human–GenAI interaction, especially through adaptive and affirming responses (Hohenstein & Jung, 2018; Hohenstein et al., 2023), may similarly fulfill emotional needs by offering safe, low-risk, and controllable social interactions.

Internet gaming researchers call this phenomenon *gamification of intimacy*, in which interactive technologies simulate emotional closeness through game-like, adaptive feedback loops. These applications blur the boundaries between fiction and lived experience, enabling *real-time narrative remediation* in which users project emotional or sexual desire onto AI entities that respond through scripted, personalized dialogue (Ge & Hu, 2025). Such affective immersion may gradually replace certain functions of human relationships. Some individuals report a decreasing desire to engage in real-world romantic or marital relationships after prolonged engagement with emotionally responsive AI systems (Zhao et al., 2025). In this context, GenAI becomes not merely a tool or companion but a substitute for emotionally complex human bonds. Based on this literature, we propose the following hypotheses:

H1: PUGenAI reflects an emotionally driven subtype of Internet gaming addiction, resembling the emotionally vulnerable type of IGD.

### Psychometric: Reuse the IGD framework in PUGenAI

To empirically assess and potentially falsify the existence of GenAI addiction, researchers must first conceptualize and operationalize what constitutes problematic use in the context of GenAI. The nine diagnostic criteria for IGD, as proposed in DSM-5, have served as a foundational framework for measuring behavioral addictions (Pontes & Griffiths, 2015) and have informed the development of instruments such as the Social Media Disorder Scale (Van Den Eijnden et al., 2016). Van den Eijnden et al. (2016) proposed that SMD and IGD may represent different manifestations of a broader category of online behavioral addiction, and therefore can be assessed using the same diagnostic framework. By extension, we argue that social media, internet games, and GenAI all serve as content in digital addiction, which suggests that PUGenAI may be examined through a similar theoretical and diagnostic lens.

Despite the emergence of several self-labeled GenAI addiction scales (Ciudad-Fernández et al., 2025), the field still lacks a systematically examined measurement. Current studies often rely on unvalidated or ad hoc instruments that have not undergone rigorous evaluation (Yankouskaya et al., 2024; Zhou & Zhang, 2024). Moreover, the cross-cultural and gender validity of AI trust measures remains largely untested. While attitudes to AI vary significantly across cultures (Barnes et al., 2024), behavioral addictions are also gender sensitive (Zakiniaeiz & Potenza, 2018), few studies examine measurement invariance. Without it, observed group differences may reflect variations in the scale's constructs rather than genuine factors of interest in inter-group studies (Putnick & Bornstein, 2016).

## Present Study

In response to recent calls for diagnostic criteria of GenAI addiction (Kooli et al., 2025) and cautions against hastily labeling GenAI use as addiction (Ciudad-Fernández et al., 2025), the present study adopts a cautious and data-driven approach (see Figure 2) through a multi-phase research program. To address this overarching aim, we propose the following four sub-research questions:

RQ1: How can the PUGenAI scale be systematically developed and validated?

RQ2: What distinct user subtypes of PUGenAI can be identified?

RQ3: What is the network structure within PUGenAI?

RQ4: Which factors predict membership in different PUGenAI user profiles?

## Phase 1

### Method

#### Item collection

We created and adapted items based on the standard DSM-5 dimensions. To avoid omission, additional items were informed by In-Depth Interview (n = 12, age Mean = 26.08, SD = 3.22; eight females) and two focus group discussions. Participant responses were collected and thematically organized into the nine dimensions (preoccupation, tolerance, withdrawal, persistence, escape, problems, deception, displacement, and conflict). As the existing DSM-5 IGD framework, the iterative coding process was guided by both inductive insights and deductive mappings to existing constructs (Proudfoot, 2022).

*Sample size*

Following Carpenter's (2018) standards, we prepared for item over-generation, creating a pool of items two to three times larger than the anticipated final item count. The ideal participant-to-variable ratio is 10:1 to 20:1. We initially generated 40 items related to PUGenAI, aiming to refine this to a 9-item scale. In the first round of data collection, we obtained over 800 samples from both the U.S. and China.

The scale development and validation require distinct samples; the second round focused on the 9-item scale. To assess criterion validity, we incorporated additional items, bringing the total to around 150. With a 10:1 participant-to-variable ratio, we estimated a sample size of 1,500, with 750 participants per group in the U.S. and China.

*Participants*

Chinese participants (aged 18 years or older) were recruited via Credamo between November 2024 and April 2025. U.S. participants (aged 18 years or older) were recruited through Prolific during the same period. In the exploratory phase, 1,508 participants completed the survey (Mean age = 34.61, SD = 11.21; 53.3% women), with 64 excluded. In the validation phase (n = 1,426, Mean age = 41.71, SD = 20.06; 47.4% women), 35 were excluded due to failed attention checks (e.g., reporting never using GenAI).

*Measurement*

Descriptions of variables are provided in Table 1. The Interaction of Person-Affect-Cognition-Execution (I-PACE) model (Brand et al., 2016) provides a comprehensive framework for explaining the development of Internet-use disorders. Drawing from dual-process theories, biopsychosocial models, and cognitive-behavioral approaches, it describes how personal predispositions interact with affective, cognitive, and executive factors to shape problematic digital behavior. Guided by this model, we organized our variables accordingly.

According to the I-PACE model, the Person component encompasses relatively stable individual characteristics, such as personality traits and psychopathological symptoms, that establish baseline vulnerability to problematic digital engagement. In this study, the following variables were included under this domain: self-construal (Gudykunst & Lee, 2003), outgroup trust (Feng et al., 2016; Torpe & Lolle, 2011), attention deficit (Kessler et al., 2005), and self-esteem (Rosenberg, 1965).

The Affect and Cognition components refer to more proximal, situationally activated states that influence users' perceived gratification and learning of technology-related behavior. These factors reinforce use patterns and shape users' emotional and evaluative responses to GenAI. Variables under this domain include loneliness (Hays & DiMatteo, 1987), rumination (Blanke et al., 2022), stress (Cohen et al., 1983), GenAI literacy (Liu et al., 2025), technology acceptance (Chin et al., 2008), and technology readiness (Lam et al., 2008).

The Execution component reflects behavioral control mechanisms, such as inhibition and decision-making. Over time, the model posits a shift from deliberate, goal-oriented use to more impulsive and habitual engagement. We operationalized this domain through frequency of daily GenAI use, differentiated by use contexts (e.g., emotional support, professional tasks, creative work, repetitive tasks), as well as perceived labor division with GenAI.

*Control and other variables*

Consistent with prior cultivation research, we measured key demographic variables, including nationality, age, gender, and education. Participants were also asked to indicate the extent to which their attitudes toward GenAI were shaped by five sources: personal usage experience, friends' opinions, internet and other media, official releases and introductions, and academic articles.

Table 1. Descriptions of Variables

| Person | Measure / Source | Scale | Sample Item / Description |
|---|---|---|---|
| Self-Construal | Self-Construal Scale (Gudykunst & Lee, 2003) | 5-point | Independent α = .919 and interdependent, α = .871 |
| Outgroup Trust | 2 items (Feng et al., 2016; Torpe & Lolle, 2011) | 7-point | "Most people can be trusted"; "Trust strangers without interest" |
| Attention Deficit | DSM-IV ADHD checklist (Kessler et al., 2005) | 5-point | "I am easily distracted", α = 0.864 |
| Self-Esteem | Rosenberg Self-Esteem Scale (Rosenberg, 1965) | 4-point | "I feel that I have a number of good qualities", α = 0.864 |
| Affect and cognition | Measure / Source | Scale | Sample Item / Description |
| Loneliness | UCLA Loneliness Scale (UCLA-8; Hays & DiMatteo, 1987) | 4-point | "I feel isolated from others", α = 0.861 |
| Rumination | 2 items (Blanke et al., 2022) | 7-point | "I could not stop thinking about my feelings." |
| Stress | Perceived Stress Scale, 4 items (Cohen et al., 1983) | 5-point | "Unable to control the important things in your life", α = 0.763 |
| GenAI Literacy | Short version of GenAI Literacy Scale (Liu et al., 2025) | 5-point | Self-perceived competence in using generative AI tools, α = .946 (total) Technical Proficiency α = .876; Critical Evaluation α = .895; Communication Proficiency α= .810; Creative Application α= .803; Ethical Competence α = .756 |
| Technology Acceptance | Technology Acceptance Scale (Chin et al., 2008) | 8-point | Perceived usability and ease of use, α = .841 (usability).α = 733 (ease) |
| Technology Readiness | Technology Readiness Index (Lam et al., 2008) | 5-point | Measures optimism, innovativeness, discomfort, and insecurity, α = .819 (total) Subdimensions α .794-.859 |
| Execution | Measure / Source | Scale | Sample Item / Description |
| Frequency of GenAI Use | Self-reported GenAI use frequency across tasks | 5-point; daily frequency (0-7) | Daily frequency + contextualized use (e.g., emotional support, creativity, professional tasks) |
| Labor Division with GenAI | Single-item evaluation | 0-100 scale | 0 = entirely human-performed; 100 = entirely GenAI-performed |

*Analysis strategy*

We simultaneously computed both alpha and omega (composite reliability) as measures of scale reliability and as references for convergent validity (Hayes & Coutts, 2020; Raykov et al., 2024). We conducted CFA using Mplus 8.3 to validate the existing scales. Following rigorous evaluation standards (Sun & Tang, 2025), we used Maximum Likelihood estimation with Robust standard errors (MLR) to estimate factor loadings and correlations and assess overall model fit (Item analysis and Discriminant validity strategy see supplemental material). Model fit with the following criteria: Comparative Fit Index (CFI) cutoff of 0.90 (Chen, 2007), Tucker-Lewis Index (TLI) cutoff of 0.90, Root Mean Square Error of Approximation (RMSEA) cutoff of 0.07 (Steiger, 2007), and Standardized Root Mean Square Residual (SRMR) cutoff of 0.08 (Hu & Bentler, 1999).

We conducted a series of nested models (Putnick & Bornstein, 2016; Vandenberg & Lance, 2000) to assess the measurement invariance of the scale across gender and nationalities. Configural invariance was established, confirming that the factor structure remains consistent across groups, implying that the same items measure the same underlying constructs. Metric invariance (weak invariance) was tested to determine if the factor loadings are equivalent across groups. Lack of metric invariance would indicate that some items may have different meanings or response biases across groups. Scalar invariance (strong invariance) assessed whether the intercepts of observed scores on latent variables are uniform across groups, suggesting that groups interpret the scale similarly. At this stage, the two factors could already be considered comparable across groups. However, we further examined whether the measurement error (residuals) for each item was invariant across groups. In addition, we assessed factor variance invariance to ensure that group differences did not affect the variability of the latent constructs. We also tested factor covariance invariance, which indicates whether the associations among latent variables remained consistent across groups. Based on the nested nature of the invariance model, changes in fit indices are compared to evaluate them (Putnick & Bornstein, 2016). A change in CFI ($\Delta$CFI) $\leq$ -0.010 and RMSEA ($\Delta$RMSEA) $\leq$ 0.015 suggests no significant deterioration in model fit, thereby supporting measurement invariance (Putnick & Bornstein, 2016).

### Result

The PUGenAI-31 model exhibited an overall excellent fit (see Supplementary Table 1), as evidenced by the CFI = 0.968, TLI = 0.957, SRMR = 0.027, and RMSEA = 0.080. In terms of discrimination, all items showed statistically significant differences between groups. All CITC values exceeding 0.45, indicating good internal consistency. The CFA of the PUGenAIS-9 model demonstrated acceptable to good model fit: CFI = 0.948, TLI = 0.931, RMSEA = 0.045, SRMR = 0.038. Inter-item correlations were moderate, indicating neither redundancy nor excessive dispersion among the nine items (see Table 2).

Table 2. Description of PUGenAIS-9 items and results of EFA and CFA

| Dimension | In the past year, have you…? |
| --- | --- |
| Preoccupation | … often found it difficult not to use GenAI when you were doing something else (e.g., school work)? |
| Tolerance | … regularly felt dissatisfied because you wanted to spend more time using GenAI? |
| Withdrawal | … often felt insecure if you couldn't access GenAI? |
| Persistence | … tried to reduce your use of GenAI, but failed? |
| Escape | … regularly used GenAI to take your mind off your problems? |

| Problems | … often got distracted at school, while doing homework, or at work, because you were thinking about using GenAI? |
| Deception | … regularly hidden your GenAI use from others? |
| Displacement | … regularly neglected other activities (e.g., hobbies, sports) because you wanted to use GenAI? |
| Conflict | … had serious problems at school or at work because you were spending too much time on GenAI? |

| Dimension | 2ndCFA FL | 1stCFA FL | Mean | SD | Skewness | Kurtosis | CITC | Cohen's $d$ |
|---|---|---|---|---|---|---|---|---|
| Preoccupation | 0.761 | 0.704 | 2.698 | 1.553 | 0.409 | -1.065 | 0.67 | 3.681 |
| Tolerance | 0.768 | 0.760 | 1.933 | 1.141 | 1.24 | 1.109 | 0.725 | 2.527 |
| Withdrawal | 0.802 | 0.836 | 2.159 | 1.355 | 0.947 | -0.166 | 0.800 | 3.436 |
| Persistence | 0.826 | 0.813 | 2.210 | 1.364 | 0.878 | -0.298 | 0.776 | 3.317 |
| Escape | 0.659 | 0.708 | 2.506 | 1.488 | 0.552 | -0.939 | 0.68 | 3.773 |
| Problems | 0.812 | 0.798 | 2.001 | 1.24 | 1.107 | 0.31 | 0.764 | 2.701 |
| Deception | 0.670 | 0.702 | 1.954 | 1.249 | 1.33 | 1.009 | 0.670 | 2.362 |
| Displacement | 0.77 | 0.767 | 1.816 | 1.114 | 1.374 | 1.232 | 0.732 | 2.279 |
| Conflict | 0.662 | 0.712 | 1.597 | 0.942 | 1.908 | 3.850 | 0.675 | 1.718 |

Note. CFA: confirmatory factor analysis, FL: factor loading, CITC: corrected item-total correlations

Correlation analyses provided evidence for the criterion validity of the PUGenAI-9 scale. As expected, PUGenAI scores showed strong positive associations with affective vulnerabilities, including rumination, stress, and loneliness. Among personal traits, ADHD symptoms were the most robust correlate, whereas self-esteem was negatively associated. In the cognitive domain, perceived discomfort, insecurity, and the importance of GenAI use were consistently linked to higher PUGenAI scores. Notably, variables such as GenAI literacy and optimism did not correlate significantly with PUGenAI. Use frequency and emotionally supportive GenAI use also emerged as significant behavioral correlates.

In cross-nationality measurement invariance, we found support for configural invariance and partial metric (loading) invariance, but full scalar invariance, residual invariance, and latent factor variances were not tenable. Thus, we proceeded with a partial invariance model by freely estimating 4 items according to the modification suggestions (see Supplementary Table 2). In cross-gender measurement invariance, changes in fit indices met recommended thresholds in all levels, even latent mean invariance (see Supplementary Table 3). These results indicate that the PUGenAI-9 functions equivalently across gender groups, with no significant difference in PUGenAI between male and female participants.

### Discussion

The PUGenAIS-9 demonstrated strong reliability and validity across two independent assessments. It also showed measurement invariance across gender and nationality, indicating its potential for broader cross-cultural application.

### Phase 2

PUGenAI reflects a form of PU, but the severity of this PU, whether it qualifies as addiction, remains uncertain. To address this, we firstly examined population-level profiles to explore (a) whether distinct symptom profiles of PUGenAI exist and (b) the prevalence of each pattern. Latent profile analysis (LPA) is a proper statistical analysis that classifies individuals into distinct symptom profiles based on their response patterns to symptoms, grouping those with similar characteristics while distinguishing them from other groups, enhancing external validity (Spurk et al., 2020). We used profile membership derived from LPA as a categorical outcome variable in a multinomial logistic regression. This allowed us to identify which cognitive, affective, and contextual indicators significantly differentiated between distinct PUGenAI symptom profiles.

In contrast, network analysis adopts a variable-centered approach by modeling the structural relationships among symptoms across the entire sample. It emphasizes how specific variables co-occur and interact, offering insights into core indicators.

### *Method*

### *Analysis strategy*

#### LPA and multinomial logistic regression

For LPA, we determined the optimal number of latent classes by evaluating multiple fit indices, including log-likelihood, Akaike Information Criterion (AIC), Bayesian Information Criterion (BIC), sample-size adjusted BIC (aBIC), and entropy. In addition, the Lo-Mendell-Rubin adjusted likelihood ratio test (LMR) has been shown to effectively identify the appropriate number of latent classes (Jenn-Yun et al., 2013). For cross-validating the robustness of profiles, we retest the LPA within the development dataset.

We checked for multicollinearity by screening for bivariate correlations and examining the Variance Inflation Index (Alin, 2010). There is no sign of multicollinearity. Given the non-ordinal nature of the profiles, multinomial logistic regression was selected as the most appropriate approach to model the probability of class membership as a function of multiple I-PACE indicators. The minimal-use group, the largest sample size, was designated as the reference category to facilitate the interpretation of odds ratios (ORs).

#### Network analysis

We used network analysis to examine the structure of PUGenAI, treating psychological variables as directly interacting elements (Borsboom, 2022). As Profiles 2, 3, and 4 exhibited symptoms of PUGenAI, they were selected for further NA, using a validation dataset (n = 1,426). To cross-validate the structural characteristics of the PUGenAI symptom network, we used the development dataset (n = 1,508) to replicate the network structure.

Following Hevey (2018), we estimated a Gaussian Graphical Model (GGM) based on 9 standardized PUGenAI items (Epskamp et al., 2018). This approach focuses on conditional dependencies among variables. We applied the EBICglasso algorithm (Hevey, 2018). To evaluate the accuracy and stability of the estimated network, we performed nonparametric bootstrapping to compute confidence intervals for edge weights. All analyses were conducted using the bootnet and qgraph packages in R.

*LPA and multinomial logistic regression Result*

Across both exploratory and verification datasets, item-level means exhibited consistent high-to-low patterns across all latent classes. In both datasets, the four-class solution was selected based on its lower information criteria and significant LMR values (Table 3).

Table 3. Fitting index of category models for latent profile analysis of PUGenAI-9

| Dataset | Number of Classes | Loglikelihood (H0) | AIC | BIC | SABIC | Entropy | LMR Adj. LRT (*p*) |
|---|---|---|---|---|---|---|---|
| verification | 3-Class | -18901.96 | 37879.92 | 38079.90 | 37959.19 | 0.909 | 0.000 |
| verification | 4-Class | -18584.39 | 37264.79 | 37517.39 | 37364.91 | 0.867 | 0.000 |
| verification | 5-Class | -18420.94 | 36957.87 | 37263.10 | 37078.86 | 0.879 | 0.398 |
| verification | 6-Class | -18279.96 | 36695.92 | 37053.78 | 36837.76 | 0.861 | 0.165 |
| exploration | 3-Class | -18470.95 | 37017.91 | 37220.01 | 37099.30 | 0.912 | 0.026 |
| exploration | 4-Class | -18076.16 | 36248.32 | 36503.61 | 36351.13 | 0.900 | 0.041 |
| exploration | 5-Class | -17921.57 | 35959.14 | 36267.62 | 36083.37 | 0.896 | 0.096 |
| exploration | 6-Class | -17772.76 | 35681.53 | 36043.19 | 35827.17 | 0.893 | 0.273 |

Note. AIC: Akaike Information Criterion; BIC: Bayesian Information Criterion; SABIC: sample-size adjusted Bayesian Information Criterion; Entropy: model classification accuracy; LMR: Lo–Mendell–Rubin adjusted likelihood ratio test.

Table 4 and Figure 3 summarize the class distribution and traits. Overall, the *Minimal Risk* profile comprised the largest group (35%–47%), characterized by the lowest levels of PU symptoms, with only escape showing minimal presence. The *Low Risk* profile (~30%) displayed slightly higher scores across all symptoms, particularly preoccupation and escape. The *Moderate Risk* profile (16%–25%) showed elevated levels of PU, with most dimensions exceeding somewhat disagree, except for tolerance, deception, and conflict. Notably, preoccupation and escape were especially pronounced. Finally, the *High Risk* group (5%–10%) reported values exceeding moderately agree across all items.

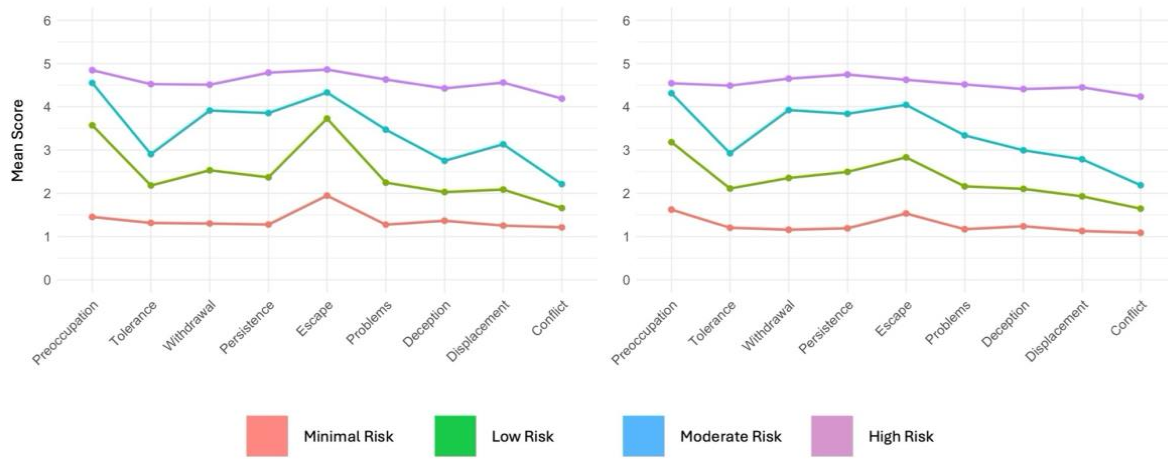

Figure 3. Distribution of the estimated conditional mean of 4 latent classes

Table 4. Profile-Level Descriptive Patterns and Predictors

| Variable | Correlation[a] | Chi-Square/$df$[b] | Class 1 (n = 491) | Class 2 (n = 140) | Class 3 (n = 359) | Class 4 (n = 436) |
|---|---|---|---|---|---|---|
| **Personal Characteristic** | | | | | | |
| US | 0.24*** | 70.16*** | 385 (78.40%) | 109 (77.90%) | 98 (27.30%) | 134 (30.70%) |
| CN | | | 106 (21.60%) | 31 (22.10%) | 261 (72.70%) | 302 (69.30%) |
| Age (years) | -0.15*** | 0.42 | 40.96 ± 12.73 | 38.09 ± 36.24 | 32.96 ± 10.83 | 34.85 ± 10.42 |
| Gender Female | 0.051 | 2.55 | 213 (43.4%) | 65 (46.40%) | 116 (32.3%) | 174 (39.9%) |
| Gender Male | | | 272 (55.40%) | 73 (52.10%) | 243 (67.70%) | 261 (59.90%) |
| Education level | 0.20*** | 5.381 | 4.70 ± 1.32 | 5.34 ± 0.99 | 5.07 ± 0.80 | 5.11 ± 0.95 |
| Independent self | -0.07** | 13.89** | 29.96 ± 6.26 | 28.45 ± 6.20 | 25.07 ± 5.04 | 25.32 ± 5.87 |
| Interdependent self | 0.06* | 3.84 | 27.30 ± 5.44 | 28.86 ± 4.57 | 25.25 ± 4.13 | 26.03 ± 4.02 |
| Self-esteem | -0.24*** | 3.88 | 48.54 ± 10.12 | 42.61 ± 8.11 | 46.67 ± 6.23 | 48.50 ± 6.12 |
| ADHD | 0.52*** | 52.54*** | 15.20 ± 6.61 | 23.27 ± 6.41 | 16.91 ± 5.03 | 14.62 ± 5.11 |
| Outgroup trust | -0.03 | 4.06 | 9.11 ± 4.01 | 9.36 ± 3.86 | 7.81 ± 2.79 | 8.15 ± 2.87 |
| **Affect** | | | | | | |
| Loneliness | 0.24*** | 2.49 | 18.50 ± 8.93 | 23.54 ± 7.48 | 18.75 ± 5.51 | 17.24 ± 5.66 |
| Rumination | 0.40*** | 22.84*** | 6.27 ± 3.63 | 10.01 ± 3.12 | 8.57 ± 3.06 | 7.07 ± 3.25 |
| Stress | 0.30*** | 8.39* | 8.82 ± 3.54 | 11.05 ± 2.63 | 9.69 ± 2.52 | 8.58 ± 2.57 |
| **Cognition** | | | | | | |
| GenAI importance | 0.31*** | 25.13*** | 4.57 ± 1.86 | 6.01 ± 0.99 | 5.69 ± 0.89 | 5.61 ± 1.05 |
| GenAI work ratio | 0.12*** | 3.48 | 54.74 ± 23.83 | 67.98 ± 18.00 | 56.10 ± 17.59 | 54.91 ± 19.23 |
| Tech readiness | -0.35*** | | 81.49 ± 12.06 | 70.73 ± 10.20 | 74.86 ± 8.21 | 78.90 ± 9.15 |
| Innovativeness | 0.03 | 8.48* | 18.18 ± 3.92 | 19.11 ± 2.96 | 16.48 ± 2.89 | 16.92 ± 3.07 |
| Optimism | -0.04 | 6.49 | 19.92 ± 3.06 | 19.61 ± 2.85 | 18.10 ± 2.23 | 18.39 ± 2.45 |
| Discomfort | 0.39*** | 31.99*** | 9.56 ± 4.39 | 14.53 ± 5.28 | 11.31 ± 3.32 | 9.67 ± 3.61 |
| Insecurity | 0.36*** | 17.54*** | 17.04 ± 7.15 | 23.46 ± 6.75 | 18.40 ± 4.89 | 16.74 ± 5.45 |
| Tech acceptance | 0.08** | | 51.89 ± 6.33 | 53.32 ± 6.13 | 52.43 ± 4.25 | 52.56 ± 4.45 |
| Usefulness | 0.09*** | 3.39 | 26.30 ± 3.79 | 27.15 ± 3.12 | 26.79 ± 2.29 | 26.82 ± 2.50 |
| Ease | 0.04 | 2.85 | 25.60 ± 3.33 | 26.17 ± 3.47 | 25.65 ± 2.57 | 25.74 ± 2.74 |
| Trust | 0.13*** | | 97.54 ± 20.51 | 104.42 ± 14.52 | 104.05 ± 12.31 | 105.87 ± 12.85 |
| Emotional Connection | 0.34*** | 23.72*** | 20.35 ± 9.88 | 28.90 ± 4.60 | 26.22 ± 4.92 | 26.20 ± 5.86 |
| Perceived Risk | 0.25*** | 8.62* | 9.81 ± 5.05 | 13.93 ± 7.16 | 11.08 ± 3.72 | 9.85 ± 3.27 |
| Competence | 0.06* | 0.52 | 29.06 ± 5.21 | 30.16 ± 3.66 | 29.55 ± 3.28 | 30.03 ± 3.37 |
| Benevolence and Integrity | 0.10*** | 1.03 | 22.95 ± 5.74 | 24.29 ± 3.57 | 24.37 ± 3.48 | 24.49 ± 3.38 |
| GenAI literacy | -0.01 | | 110.19 ± 19.03 | 114.25 ± 16.19 | 106.95 ± 14.96 | 110.52 ± 15.61 |
| Technical Proficiency | 0.07** | 3.04 | 25.54 ± 6.42 | 28.07 ± 4.65 | 25.82 ± 4.52 | 26.71 ± 5.07 |
| Critical Evaluation | -0.06* | 12.49** | 31.29 ± 6.30 | 32.04 ± 5.35 | 29.68 ± 5.21 | 30.71 ± 5.10 |
| Communication Proficiency | -0.09*** | 7.56 | 21.25 ± 3.42 | 20.76 ± 3.10 | 20.15 ± 2.86 | 20.82 ± 2.82 |
| Creative Application | 0.06* | 0.44 | 16.18 ± 3.62 | 16.96 ± 2.50 | 16.27 ± 2.45 | 16.68 ± 2.44 |
| Ethical Competence | -0.04 | 9.43* | 15.92 ± 3.32 | 16.41 ± 2.57 | 15.03 ± 2.82 | 15.59 ± 2.80 |

| Execution | | | | | | |
|---|---|---|---|---|---|---|
| Daily GenAI use frequency | 0.12*** | 6.94 | 4.32 ± 1.70 | 5.58 ± 1.50 | 4.14 ± 1.58 | 4.21 ± 1.58 |
| Emotion support use | 0.27*** | 3.30 | 2.48 ± 1.33 | 3.74 ± 1.00 | 2.91 ± 1.10 | 3.01 ± 1.13 |
| Creative use | 0.23*** | 0.69 | 3.02 ± 1.32 | 3.94 ± 0.85 | 3.62 ± 0.92 | 3.62 ± 1.02 |
| Repetitive task use | 0.22*** | 4.46 | 3.03 ± 1.31 | 3.89 ± 0.83 | 3.69 ± 0.95 | 3.72 ± 1.00 |
| Professional task use | 0.19*** | 0.73 | 3.04 ± 1.25 | 3.88 ± 0.86 | 3.46 ± 0.95 | 3.42 ± 1.09 |
| Others | | | | | | |
| PUGenAI score | | | 12.40 ± 3.10 | 41.31 ± 4.27 | 31.09 ± 2.96 | 22.40 ± 2.81 |
| Personal experience | -0.04 | 0.92 | 4.50 ± 0.70 | 4.39 ± 0.67 | 4.27 ± 0.66 | 4.29 ± 0.64 |
| Friends' opinion | 0.25*** | 2.44 | 2.95 ± 1.33 | 3.87 ± 0.83 | 3.42 ± 0.81 | 3.37 ± 0.87 |
| Online media | 0.16*** | 1.46 | 3.60 ± 1.12 | 4.06 ± 0.75 | 3.78 ± 0.80 | 3.75 ± 0.81 |
| Official source | 0.19*** | 0.96 | 3.28 ± 1.31 | 3.94 ± 0.89 | 3.65 ± 0.95 | 3.69 ± 0.95 |
| Academic source | 0.22*** | 12.21*** | 3.15 ± 1.35 | 4.07 ± 0.87 | 3.53 ± 1.00 | 3.66 ± 1.06 |

Note. a. Correlations represent the results of examining external validity, using the sum score of PU to correlate with other factors. b. Chi-Square/*df* represents the results of the multinomial logistic regression. The significance level indicates whether a factor significantly predicts membership in the four PU profiles.

The final multinomial logistic regression model demonstrated a significantly improved fit over the intercept-only model, $\chi^2 (117) = 1232.97$, $p < .001$, Nagelkerke $R^2 = .627$. Guided by the correlation and multinomial logistic regression results, we employed the I-PACE model as an integrative framework to organize the robust predictors of PUGenAI. Individuals identified as problematic users exhibited significantly higher scores on ADHD (high: OR = 1.206, $p < .001$, moderate: OR = 1.185, $p < .001$, and low: OR = 1.077, $p < .001$), and independent self-construal (high: OR = 0.881, $p < .001$). Additionally, nationality was associated with problematic use (high: OR = 0.064, $p < .001$, moderate: OR = 0.026, $p < .001$, and low: OR = 0.097, $p < .001$).

Problematic users also scored higher on affective variables, including rumination (high: OR = 1.190, $p = .002$, moderate: OR = 1.189, $p < .001$, and low: OR = 1.104, $p = .003$), and stress (high: OR = 1.231, $p = .005$). On the cognitive level, in technology readiness, perceived insecurity dimension (high: OR = 1.109, $p < .001$, and moderate: OR = 1.066, $p = .002$), discomfort (high: OR = 1.195, $p < .001$, moderate: OR = 1.151, $p < .001$, and low: OR = 1.057, $p = .035$), and innovativeness dimension (high: OR = 1.224, $p = .009$) showed significant associations. However, GenAI literacy was not positively associated with problematic use. Specifically, critical evaluation was negatively associated in the moderate (OR = 0.919, $p = .005$) and low (OR = 0.922, $p = .002$) groups. Similarly, ethical competence was negatively associated with problematic use across all levels (high: OR = 0.846, $p = .029$, moderate: OR = 0.868, $p = .004$, and low: OR = 0.910, $p = .027$). We observed a misalignment across trust dimensions. Emotional connection (high: OR = 1.209, $p < .001$, moderate: OR = 1.092, $p = .001$) and perceived risk (high: OR = 1.105, $p = .017$) were elevated among problematic users, whereas competence-based trust and benevolence and integrity were notably weaker.

*Network Analysis Result*

In the networks, tolerance, withdrawal, persistence, problems, and conflict consistently emerged as highly central nodes across both networks (Table 5). These nodes demonstrated high strength, closeness, and betweenness centrality, suggesting that these five dimensions play a pivotal role in sustaining and coordinating the problematic use behavior. The escape and preoccupation consistently occupied a peripheral position, with low closeness and betweenness scores in both networks, indicating its limited connectivity to other nodes (Figure 4A and 4B and Table 3). Figure 4 C and D illustrate the accuracy of edge-weight estimates using 95% nonparametric bootstrap confidence intervals. Most edges demonstrated acceptable stability, with confidence intervals that did not include zero, suggesting these associations are reliably estimated.

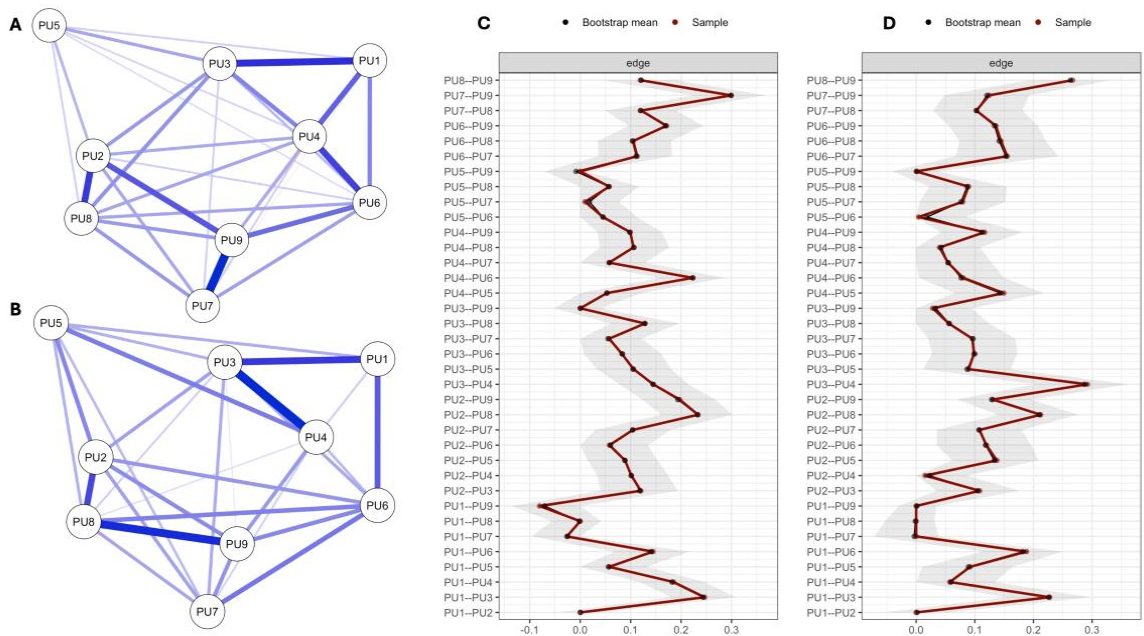

Figure 4. Symptom Network Structures and Edge Accuracy

Note: Panels A and B depict the estimated symptom network structures of PUGenAI in the validation and development datasets, respectively. The layout in Panel B replicates the configuration derived from Panel A to ensure visual comparability. Blue lines represent positive associations, and red lines represent negative associations. The thickness and saturation of each edge reflect the strength of the association. Nodes that are more strongly or centrally connected appear closer together, with the most central symptoms positioned near the center of the network. Panels C and D display the edge-weight accuracy plots for the validation and development datasets, respectively, estimated using 95% nonparametric bootstrap confidence intervals. Narrower confidence intervals indicate more stable edge estimates. Edges with intervals that include zero should be interpreted with caution, as they may reflect weak or unreliable associations.

Table 5. Comparison of Node Centrality Metrics Between Networks

| | Network A | | | | Network B | | | |
|---|---|---|---|---|---|---|---|---|
| Node | Betweenness | Closeness | Strength | Expected Influence | Betweenness | Closeness | Strength | Expected Influence |
| PU1 (Preoccupation) | 2 | 0.012 | 0.740 | 0.526 | 1 | 0.013 | 0.565 | 0.565 |
| PU2 (Tolerance) | 4 | 0.014 | 0.903 | 0.903 | 3 | 0.014 | 0.832 | 0.832 |
| PU3 (Withdrawal) | 5 | 0.013 | 0.885 | 0.885 | 4 | 0.014 | 0.996 | 0.996 |
| PU4 (Persistence) | 0 | 0.013 | 0.969 | 0.969 | 2 | 0.013 | 0.805 | 0.805 |
| PU5 (Escape) | 0 | 0.008 | 0.417 | 0.417 | 0 | 0.011 | 0.641 | 0.641 |
| PU6 (Problems) | 1 | 0.013 | 0.942 | 0.942 | 3 | 0.014 | 0.926 | 0.926 |
| PU7 (Deception) | 0 | 0.011 | 0.788 | 0.735 | 0 | 0.012 | 0.720 | 0.720 |
| PU8 (Displacement) | 1 | 0.013 | 0.870 | 0.870 | 0 | 0.013 | 0.911 | 0.911 |
| PU9 (Conflict) | 4 | 0.013 | 0.967 | 0.805 | 1 | 0.013 | 0.803 | 0.803 |

Note. Network A presents the results based on the validation dataset. Network B illustrates results based on the exploration data.

## Discussion

The current study identified four distinct latent profiles of problematic GenAI use, ranging from minimal use to high addiction symptom profiles. Despite clear gradations in overall symptom severity, the symptom configuration remained highly consistent across classes, suggesting a unidimensional rather than multidimensional structure of PUGenAI. Further inspection of the item means indicated that preoccupation, withdrawal, and escape symptoms were consistently rated highest across classes, while tolerance and conflict were rated lowest. This pattern reflects the prevalence of affect-regulation motives in GenAI engagement. Conversely, the consistently lower symptoms indicate that GenAI use may induce psychological salience without yet manifesting in social or behavioral dysfunction for most users. Results of logistic regression suggest that PUGenAI may be driven more by emotional than by instrumental evaluation of system capabilities.

Both escape and preoccupation consistently ranked lowest in network centrality indices. This apparent discrepancy suggests that while these motives are commonly endorsed, they do not play a structurally integrative role within the network of problematic GenAI use. Instead, they may reflect a generalized surface-level cause. This pattern aligns with the emotionally vulnerable IGD, in which preoccupation and escape function more as entry points than as structural cores of the addiction network (Lee et al., 2017). Moreover, a latent structure previously observed in IGD was replicated, with conflict, withdrawal, and tolerance emerging as the most central symptoms (Gomez et al., 2022; Liu et al., 2022).

## General Discussion

The present study draws upon the nine diagnostic criteria for IGD outlined in the DSM-5 (American Psychiatric Association, 2013) to develop a measurement instrument tailored to GenAI use. The study aims to evaluate the applicability of this addiction diagnostic framework within the GenAI context. Rather than prematurely asserting a formal clinical category, we employed both top-down (DSM-based) and bottom-up (data-driven) strategies to construct the PUGenAI scale. Empirical analyses were conducted across two phases: (a) structural validation of the scale, (b) identification of symptoms. Findings revealed that PUGenAI is best understood as an affect-driven, emotionally vulnerable subtype of IGD, characterized by psychological vulnerability, emotional reward dependence, and weakened critical literacy rather than by functional reliance or skill deficits. This finding responds to ongoing debates regarding the blurred boundaries between addiction, dependence, and problematic use, as well as the lack of consensus around measurement standards in behavioral addiction research.

### Affective Vulnerability as the Core of PUGenAI

The emotionally vulnerable subtype of IGD is characterized by elevated preoccupation and escape tendencies (Lee et al., 2017). Consistent with this, the LPA identified three PUGenAI risky symptom profiles that exhibited elevated scores in preoccupation and escape, suggesting these symptoms are particularly salient in situationally triggered symptom profiles. This aligns with the finding that emotional connection serves as a robust indicator of PUGenAI. Network analysis further corroborates this pattern: across both exploratory and confirmatory datasets, preoccupation and escape emerged with the lowest centrality values in all three PUGenAI profiles, suggesting that affective indicators may serve as the foundational conditions for problematic engagement. Moreover, functional literacy dimensions, such as creative application, communication proficiency, and technical proficiency, did not significantly predict PUGenAI profiles. In contrast, critical evaluation (e.g., "I can determine whether GenAI's response is true") and ethical competence (e.g., "I can recognize potential privacy issues related to the use of GenAI") negatively predicted PUGenAI, which highlights the importance of critical thinking rather than skill-based characters in Human–AI interaction (Laffier et al., 2025).

Drawing on these findings, we conceptualize PUGenAI as a behaviorally dysregulated state characterized by psychological vulnerability, affective reward dependence, and the erosion of cognitive literacy mechanisms. This supports H1, which posits that PUGenAI resembles an affect-driven subtype of IGD. However, competence-based trust did not significantly predict any PUGenAI profile; this is consistent with prior research emphasizing the transformative role of anthropomorphized AI agents in forming Human–AI intimacy (Hu et al., 2025). Related findings suggest that escape and social motivations mediate the link between mental health challenges and AI dependence, whereas entertainment and instrumental motivations do not (Huang et al., 2024).

These conclusions also indicate gamified intimacy, fostering affective connections and user engagement (Ge & Hu, 2025). Similarly, studies in virtual relationship contexts reveal that most participants perceive their second-life relationships as authentic rather than playful. They expect idealized traits in virtual partners, experience immersion in these interactions, and report impacts on offline relationships (Gilbert et al., 2011).

### Empirical Evidence for the Addictive Potential of PUGenAI

Although we adopt the term *problematic use* to avoid pathologizing high-frequency users, we do not discount the addiction-like qualities of PUGenAI. Our findings suggest that the PUGenAI exhibits considerable pathological or addiction-related potential. Four lines of evidence support this claim: prevalence patterns, symptom network structure, predictive correlates, and its conceptual alignment with the theoretical continuum perspective.

First, behavioral addictions are defined by impaired control, negative consequences, psychological distress, and functional impairment (Billieux et al., 2017; Kardefelt-Winther et al., 2017). Following this definition, we developed the PUGenAIS-9 scale based on the DSM-5 criteria for IGD. Notably, 5% to 10% of participants consistently scored high across all nine items, a prevalence pattern comparable to those observed in IGD studies (APA, 2013; Liao et al., 2022; Stevens et al., 2020), indicating the presence of a substantial at-risk group.

Second, findings from both network analysis and LPA strengthen the clinical plausibility of PUGenAI. Specifically, preoccupation and escape, core symptoms of the emotionally vulnerable IGD subtype (Lee et al., 2017), exhibited high mean scores but low centrality across profiles. This pattern suggests that while these symptoms may act as common initiators of problematic use, they do not function as structural anchors. In contrast, other symptoms associated with behavioral addiction, such as tolerance, withdrawal, problems, and conflict, consistently demonstrated high centrality across both samples. These findings mirror the latent structure previously observed in IGD (Gomez et al., 2022; Liu et al., 2022), reinforcing the validity of a PUGenAI subtype.

Robust predictors of PUGenAI severity encompass both psychological distress and functional impairment, two core criteria for clinical diagnosis (Billieux et al., 2017; Kardefelt-Winther et al., 2017). These are reflected in strong associations with cognitive risk factors (e.g., perceived risk, discomfort, and insecurity), consistent with prior IGD research showing elevated frustration discomfort among individuals with IGD (Lin et al., 2021). Affective vulnerabilities such as stress and rumination also emerged as strong predictors of PUGenAI, an indicator often linked to IGD (Kaess et al., 2017). Individual-level predictors, including low independent self-construal and high ADHD symptoms, further underscore the role of personal vulnerability in compulsive engagement. Cross-national differences also highlight the influence of sociocultural factors: the prevalence of gaming disorder reaches 12% in East Asia (Liao et al., 2022), compared to 3.05% globally (Stevens et al., 2020). These discrepancies may reflect broader cultural dynamics, particularly the interplay between individualism-collectivism orientations and attitudes toward both AI and the prevalence of IGD (Barnes et al., 2024; Stavropoulos et al., 2021). Interestingly, in contrast to IGD research showing a higher prevalence among males than females (16% vs. 8%; Liao et al., 2022), gender did not emerge as a robust predictor of PUGenAI. However, ADHD, as a transdiagnostic predictor of IGD (Yen et al., 2017), remained positively correlated with higher PUGenAI.

Notably, the total score of digital literacy showed no significant association with PUGenAI across analyses. These findings challenge assumptions that problematic use arises from a lack of cognitive skill. Instead, it suggests that compulsive use is driven more by affective and motivational mechanisms than by deficits in knowledge. This interpretation aligns with previous findings on AI addiction, where individuals with strong autonomy or relatedness needs exhibited increased dependence and difficulty disengaging from AI under high perceived risk (Salah et al., 2024).

Finally, consistent with theoretical positions advanced in the DSM-III and DSM-IV revisions, dysregulated behavior should be understood as a continuum rather than as a binary diagnostic threshold (Widiger & Smith, 1994). From this spectrum-based perspective, emotionally driven PUGenAI may be conceptualized as a variant of IGD.

As affective support from AI can serve as a constructive coping strategy for relatively healthy individuals (Brandtzaeg et al., 2025), caution is warranted when interpreting emotional reliance as inherently pathological. Research has shown that online friendships and romantic relationships often mirror the emotional depth and functional significance of their offline counterparts. Individuals frequently spend more time communicating with friends met online than with those known offline (Coulson et al., 2018). Therefore, diagnosing emotional dependence as problematic should not be based solely on modality or frequency of interaction (Lee et al., 2017). However, emotional reliance may signal maladaptive use when individuals are unable to regulate mood through alternative healthy strategies and become excessively dependent on AI-based experiences for emotional relief. In such cases, escapism can be interpreted as a dysfunctional coping mechanism, more likely indicative of disordered behavior, such as IGD (Lee et al., 2017).

### *Rethinking the Relationship Between Technology and Addiction*

This study identifies addiction-like symptoms in PUGenAI, yet we argue that such symptoms represent an emotionally vulnerable IGD. As technologies evolve, behavioral addictions increasingly overlap across platforms and modalities. PUGenAI often involves the technical infrastructure of the internet and the hardware affordances of smartphones, yet no study to date has systematically conceptualized the distinctions or overlaps among different forms of digital addiction. Considering this, we are against prematurely labeling new forms of addiction. Digital addictions may be more accurately understood with ICD (infrastructures, devices, and content) model (see Figure 5), distinguishing among: (1) infrastructure (e.g., the internet, large language models); (2) content-based objects (e.g., watching shows, gaming, social media); and (3) Devices (e.g., television, smartphones, computers). Infrastructure is invisible but embedded in everyday practice, like the Internet (Star & Ruhleder, 1996). Smartphones or computers serve as the physical interface, while addictive behavior typically targets specific content domains, such as GenAI, gaming, or social media.

Figure 5. ICD Model of Digital Addiction

| Dimension | Definition | Term Examples | Addiction Attribution |
|---|---|---|---|
| Infrastructure | The underlying systems that enable content functionality and access | The Internet addiction | An enabling condition, but rarely the object of addiction itself, given its transparency. |
| Content | The specific content or activity that elicits compulsive engagement | Internet gaming addiction, gambling addiction | The primary target of addiction; content drives compulsive use |
| Device | The hardware or device through which media is accessed | Smartphones addiction, computer addiction | Not the core object of addiction; only the medium of delivery |

In the case of gambling disorder conducted on smartphones, the addiction should be attributed to gambling content, not the computer or the Internet. This perspective emphasizes that addiction is content-oriented at its core, and conceptual clarity is required to differentiate symptoms stemming from shared neural mechanisms versus those shaped by affordances. This model may explain the high correlation between PUGenAI and smartphone dependency (Ali et al., 2025), as the behavioral target often lies in the emotionally gratifying content rather than the technology itself. Therefore, in addressing problematic use as shaped by the affordances of emerging technologies, researchers should critically reflect on the core mechanisms underlying such behaviors. Integrating empirical evidence into the proposed ICD model can help clarify conceptual boundaries, prevent the proliferation of loosely defined terms, and re-anchor the term in a framework that is connected to psychiatric and clinical paradigms.

## Limitations and future study

We acknowledge two key limitations of this research. First, more empirical work is needed to examine whether PUGenAI reflects distinct neurocognitive patterns, such as activation in emotion regulation or reward-related brain regions. Physiological evidence could strengthen or refute the claim that GenAI overuse engages the same circuits implicated in behavioral addictions.

Second, most existing research (including our own) is cross-sectional. However, longitudinal studies are essential for clarifying whether PUGenAI causes sustained dysfunction or reflects context-dependent dysregulation. Preliminary evidence suggests that only mental health problems predict later AI dependence, not vice versa (Huang et al., 2024). The potential for spontaneous recovery (Billieux et al., 2015) should be investigated to avoid mislabeling adaptive coping as pathological behavior.

## Conclusion

The field of behavioral addiction research currently faces a conceptual divergence. One trajectory focuses on the mechanisms of new technology addiction. The other path remains rooted in the ongoing definitional debate. Our study attempts to strike a balance. This study examined whether addiction to GenAI, when assessed using DSM-5 criteria, reflects clinical significance. Using samples from both China and the United States, we developed the PUGenAI 9-item scale. Measurement invariance across nationality and gender demonstrated that the PUGenAI scale performs consistently, supporting its applicability for cross-national research.

External validity analyses, latent profile modeling, network structure, and logistic regression all converge on a key conclusion: PUGenAI exhibits addiction-like characteristics, particularly those aligned with the emotionally vulnerable subtype of IGD, rather than competence-based dependency. Considering the overlap between digital addictions, we call for a reframing of digital addiction research through an ICD model that differentiates among infrastructure, content-based objects, and devices. This approach offers a theoretically grounded framework for interpreting problematic media use, one that resists over-pathologization and destigmatizes emerging forms of technological engagement.

## Footnote

The Diagnostic and Statistical Manual of Mental Disorders (DSM), first published by the American Psychiatric Association, has undergone several revisions, from DSM-III (1980) to DSM-5 (2013).